\documentclass[twocolumn,english,superscriptaddress,aps,prl,showpacs,amsmath,amssymb]{revtex4}
\usepackage[T1]{fontenc}
\usepackage[latin1]{inputenc}
\usepackage{amsmath}
\usepackage{graphicx}
\usepackage{amssymb}

\makeatletter

\providecommand{\LyX}{L\kern-.1667em\lower.25em\hbox{Y}\kern-.125emX\@}

\usepackage{graphicx}
\usepackage{subfigure}
\usepackage{graphicx}
\newcommand{\be}{\begin{equation}}
\newcommand{\ee}{\end{equation}}
\newcommand{\bea}{\begin{eqnarray}}
\newcommand{\eea}{\end{eqnarray}}

\usepackage{graphicx}
\usepackage{dcolumn}
\usepackage{bm}

\usepackage{babel}
\makeatother
\begin{document}

\title{Dynamic roughening and fluctuations of dipolar chains}

\author{Renaud Toussaint}

\affiliation{Department of Physics, University of Oslo, P.O. Box 1048 Blindern,
N-0316 Oslo, Norway}

\email{Renaud.Toussaint@fys.uio.no}

\author{Geir Helgesen}

\affiliation{Department of Physics, Institute for Energy Technology, N-2027 Kjeller, Norway}

\author{Eirik G. Flekk\o y}

\affiliation{Department of Physics, University of Oslo, P.O. Box 1048 Blindern,
N-0316 Oslo, Norway}

\date{\today{}}

\begin{abstract}
Nonmagnetic particles in a carrier ferrofluid acquire an effective
dipolar moment when placed in an external magnetic field. This 
fact leads them to form chains that will roughen due to Brownian 
motion when the   magnetic field is decreased.
We study this process through experiments, theory and simulations,
three methods that agree on the scaling behavior over 5 orders of
magnitude.  The RMS width goes initially as 
$t^{1/2}$, then as $t^{1/4}$ before it saturates.
We show how these results complement existing results on polymer
chains, and how the chain dynamics may be described by a recent
non-Markovian formulation of anomalous diffusion.

\end{abstract}



\maketitle
Magnetic holes \cite{Skj83,Toussaint03,SH91} are nonmagnetic micrometre 
sized 
spheres suspended in a ferrofluid, much larger than the magnetites
in suspension (nm). In an external magnetic field $\mathbf{H}$, these
holes acquire an effective dipolar moment equal to the opposite of
the dipolar moment of the displaced ferrofluid. This is true for any
non-magnetic material that makes up the spheres. When placed between
two non-magnetic glass plates the spheres aquire interactions that
may be fine tuned to produce well defined separation distances and
inter particle forces. We have recently obtained the analytic form
of these interactions and verified it experimentally \cite{Toussaint03}.
Among the many intriguing structures that may be predicted from this
theory are particle chains.

Collective Brownian motion in particle chains have been studied extensively
\cite{Furst00} over the last decade. In general this motion is governed
by a complex interplay of different mechanisms such as hydrodynamic
particle--particle interactions, particle and fluid inertia, chain-chain
interactions, Brownian forces and direct viscous drag forces. The
magnetic hole chains represent a simple prototype system that is designed
to eliminate all but the essential mechanisms needed to produce non-trivial
collective behavior. The isolation of these mechanisms, which are
Brownian fluctuations, viscous drag and an anisotropic interparticle
attraction, allows a straightforward theoretical treatment of the kinetic
roughening process. These predictions are confirmed both by our experiments
and simulations over 5 orders of magnitude in the dynamic domain.
The size of this domain along with the theoretical simplicity of the
model appears to allow significantly more conclusive statements both
on the scaling behavior of the chain motion and on the associated
prefactors, than in existing works.

By studying dynamic roughening of initially straight chains we make
contact with the theory of kinetic interface growth processes. It
is shown that since the long range hydrodynamic interactions are eliminated
in our setup by the presence of confining interfaces at a distance
shorter than the chain length, the continuum description of the chain
roughening process conforms to the Edwards-Wilkinson equation \cite{edwards82}.
The model developped here is also a discrete generalization of a Rouse
model \cite{Grosberg94}, first developped to study polymer dynamics.

In the end we will  show how the Markovian N-particle description 
of the entire chain may be be contracted to a non-Markovian description
of a single particle in the chain. This is done by integrating out all
interaction degrees  of freedom. 
What is left is a generalized Langevin equation with long term 
memory. It has recently been  shown how such an equation may be used 
to predict anomalous diffusion exponents  \cite{MOBH02}, and indeed
these exponents coincide with our independent predictions and measurements.
\begin{figure}
\includegraphics[  scale=0.40,
  keepaspectratio]{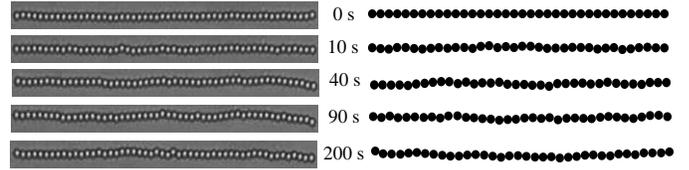}

\caption{\label{fig1}Typical dynamic roughening of dipolar chains of 52 magnetic
holes of diameter $4\mu m$ after a sudden decrease of the external magnetic
field.  Experiments are 
shown in gray and simulations black and white.}
\end{figure}

The ability of particle chains to change the rheological properties
of their carrier fluids has given rise to  practical
applications and designs such as dampers, hydraulic valves, clutches
and brakes \cite{Bullough96}. In constant fields (electric or magnetic,
depending on the nature of the dipoles), these chains aggregate laterally
\cite{Halsey90} due to their thermal fluctuations. Therefore, the
precise quantification and understanding of these fluctuations along an
\textit{isolated} dipolar chain is an important component
to understand the aggregation phenomena in constant fields of MR \cite{Martin99}
or ER \cite{Fraden89} fluids, as well as in systems of magnetic holes
\cite{Helgesen03}.

We note that also by coating the microspheres with bio-active materials,
such as streptavidin \cite{wirtz95}, they may be used for the direct
manipulation of single strands of DNA. For such application the quantitative
control of Brownian fluctuations is essential.

In the experiments monodisperse polystyrene spheres \cite{Uge80}
of diameters $a=$ 3.0 or 4.0 $\mu $m were dispersed in a kerosene
based ferrofluid \cite{Fer} of susceptibility $\chi =0.8$ and viscosity
$\eta =6\cdot 10^{-3}\mbox{Pa}\; \mbox{s}$, inside a glass cell of size 38 mm
$\times $ 8 mm $\times $ 10 $\mu $m. 
A pair of  outside coils produced magnetic field strengths up to $H=$ 20
Oe.  The setup was mounted under an
optical microscope with an attached video camera recording four frames
per second.  Low volume fractions (< 1\%)
of microspheres were used and chains were grown \cite{Helgesen03} 
by applying a constant field
of about $H=$ 18 Oe parallel to the thin ferrofluid layer for about
20 minutes. The cell was searched for long isolated chains of 30-120 spheres.
 The field was then reduced to a
constant value $H$ in the range $2$ Oe $\leq H\leq 10$ Oe while
the motion of one long chain was recorded and analyzed. 
One pixel of the video image corresponded to
$0.5 \mu$m, and the uncertainty in particle
position could be reduced to  0.2 $\mu $m by utilizing the intensity profile of the pixels showing the particle. The 
experiments illustrated in  Fig.~\ref{fig1} are challenging in part 
because this 
accuracy is needed to reveal the scaling behavior of the chains.

In order to obtain an equation of motion for chains we define
 the  lateral displacement of particle $i$,   $h_{i}$,  from  the initial 
straight line. A particle pair  at a separation $r$ and 
 angle $\theta $ to  the external
magnetic field, experiences a coupling energy 
$U=\mu _{f}\sigma ^{2}(1-3\cos ^{2}\theta )/(4\pi r^{3})$ 
where  $\sigma =-V\bar{\chi }\mathbf{H}$
\cite{Skj83,BB78}, and  $V$ is the hole's volume and $\bar{\chi }=3\chi /(3+2\chi )$
takes into account the demagnetization factor of spherical particles.
 The ratio of the maximum interaction energy over the
thermal energy $k_{B}T$ characteristic of the random forces due to
the molecular motion in the solvent is defined as \cite{DeGennes70}
$\lambda =(\mu _{f}\sigma ^{2})/(2\pi a^{3}kT)$.
In Fig.~\ref{fig1} the initial 
value $\lambda =866$ is reduced to $\lambda =24$.
Neglecting all magnetic interactions but the nearest neighbor ones
and performing a Taylor expansion of the magnetic interaction potential
around the minimal energy configuration, a straight line with spacing
$a$, the lateral component of the magnetic force on sphere $i$ is
$F_{i}^{M}=\alpha (h_{i+1}-2h_{i}+h_{i-1})$, with $\alpha =\pi \mu _{f}{\bar{\chi }}^{2}H^{2}a/12=6\lambda \cdot kT/a^{2}$.
Since the Reynolds number in this system is very small (typically
Re$=10^{-5}$), the hydrodynamic forces are linear in the particle
velocity, and $F_{i}^{H}=-\kappa \dot{h}_{i}$ where $\kappa =3\pi \eta a$.

Newton's second law for the $i$-th sphere is then
 \begin{equation}
m\ddot{h}_{i}=F_{i}^{M}+F_{i}^{H}+\zeta _{i}(t)\label{first}
\end{equation}
 where the fluctuating force $\zeta _{i}(t)$ is due to the molecular
nature of the fluid and gives rise to the Brownian motion of the particle.
At time scales exceeding the viscous damping 
time $t_{m}=m/\kappa =a^{2}\rho /18\eta \approx 10^{-7}$
s, the inertial term $m\ddot{h}_{i}$ is negligible.
 Due to the presence of confining
plates the inertial motion in the fluid also decays 
on this time scale, so that we can neglect any non-Markovian
corrections to the above equations as well as 
such corrections  in the fluctuating force
\cite{flekkoy96d}.
 We may therefore write
 $\langle \zeta _{i}(t)\zeta _{j}(0)\rangle =2\kappa k_{B}T\delta (t)\delta _{ij}$,
where the prefactor reflects the
equipartition of particle kinetic energy, i.e.
 ${k_{B}T}=\langle m\dot{h}_{i}^{2}\rangle $ \cite{reif65}.
Combining the above equations Eq.~(\ref{first}) can  be written 
\begin{equation}
\dot{h}_{i}=\frac{\alpha }{\kappa }(h_{i+1}+h_{i-1}-2h_{i})+\frac{1}{\kappa
}\zeta _{i}(t)\; .\label{discrete}
\end{equation}
 For spatial scales above $a$ and times above $t_{m}$, the above
reduces to the Edwards--Wilkinson equation \cite{edwards82} $\kappa \partial h/\partial t=\alpha a^{2}\partial ^{2}h/\partial ^{2}x^{2}+\zeta _{i}(t)\; ,$
also known in polymer dynamics as the Rouse model \cite{Grosberg94}.

We consider an isolated chain of $N$ particles, and are interested
in the dynamic roughening of the chain. To observe 
this experimentally $\lambda $ is decreased
 from a value $\lambda _{0}\gg 1$, to a finite value
still greater than $1$ (to ensure that the chain does not melt),
and the root mean square width of the displacements
along the chain is recorded.

 It is convenient to describe this by 
using the discrete space-Fourier transform along
the chain $\tilde{h}_{n}=\frac{1}{N}\sum _{j=0}^{N}h_{j}e^{-2i\pi nj/N}$,
for $n=0...N-1$. 
Equation (\ref{discrete})  then takes the form
\begin{equation}
\dot{\tilde{h}}_{n}=-\omega _{n}\tilde{h}_{n}+\tilde{\zeta }_{n}/\kappa \, \, ,\label{h}\end{equation}
 with  the dispersion relation $\omega _{n}=2\alpha (1-\cos (2\pi n/N))/\kappa \; ,$
and random terms obeying $\langle \tilde{\zeta }_{m}(t)\tilde{\zeta }_{n}^{*}(0)\rangle =2\kappa k_{B}T\delta (t)\delta _{mn}/N\; .$
Equation (\ref{h}) is easily solved to give \begin{equation}
\tilde{h}_{n}(t)=\tilde{h}_{n}(0)e^{-\omega _{n}t}+\int _{0}^{t}dt'\;
 e^{-\omega _{n}(t-t')}\tilde{\zeta }_{n}(t')/\kappa \, \, .
\label{eq:rel,h-n,zeta-n}
\end{equation}
Setting $\tilde{h}_{n}(0)=0$ and taking the thermodynamic average of the square
of the above equation leads to
\begin{equation}
\langle \tilde{h}_{n}(t)\tilde{h}_{m}^{*}(t)\rangle =\frac{k_B T}{\kappa
  N}\, \frac{1-e^{-2\omega _{n}t}}{\omega _{n}}\, \delta _{mn}\,
.\label{eq:square;h-n;initial;straight}
\end{equation}
 for $n\neq 0$, and$\langle |\tilde{h}_{0}(t)|^{2} \rangle 
=(2k_{B}T/\kappa N)t$.
It is seen from Eq.~(\ref{eq:square;h-n;initial;straight})
that each Fourier mode is initially in a free diffusion regime,
$\langle |\tilde{h}_{n}(t)|^{2}\rangle \sim  t$,
for  $t \ll \tau _{n}=1/2\omega _{n}$ and saturates  when $t \gg \tau_{n}=1/2\omega _{n}$.
The minimum and maximum saturation
times are respectively $\tau _{N/2}=\tau =\kappa /8\alpha $
and  $\tau _{1}=N^{2}\tau /\pi ^{2}$ for the shortest and longest
wavelength. 

\begin{figure}
\includegraphics[  width=0.80\columnwidth,
  height=0.80\columnwidth,
  keepaspectratio,
  angle=270,
  origin=c]{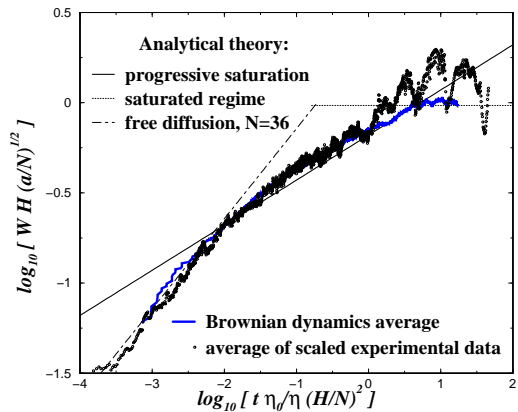}
\caption{\label{fig3}Scaling of the random mean square width of the transverse
displacements: theory, experiments and brownian dynamics result.
The units are Oe, $\mu$m and seconds for  $H$, $a$, $W$ and $t$ respectively.}
\end{figure}

We are interested in 
the mean square width of the chain 
$W^{2}=\sum _{i}(h_{i}-\sum _{j}h_{j}/N)^{2}/N=\sum _{n=1}^{N-1}\langle |\tilde{h}_{n}|^{2}\rangle$.
When $t\ll \tau $ 
all Fourier modes are in free diffusion and $W^{2}\approx 2k_{B}Tt/\kappa \; .$
This result arises only because of the existence of a shortest wavelength
$a$ in the system. 
In the continuum limit 
$a\rightarrow 0$, $\tau \rightarrow 0$ and this regime does not exist. 
Later on the modes associated with progressively 
longer wavelengths reach their 
saturated states and this is reflected in a
new  scaling behavior of $W$.
By inserting Eq.~(\ref{eq:square;h-n;initial;straight}) in the above
expression for $W^2$ it is straightforward to show that it satisfies
the Family-Vicsek scaling form $W^{2}=NF\left(t/N^{2}\right)$.
Moreover, the exact form of $F$ may be obtained.
Along with  the expressions for $\tau $, $\kappa $ and $\alpha $
this gives the result
\[
\frac{W^{2}H^{2}a}{Nk_{B}T}=\! \left\{ \! \begin{array}{lr}
 \! \frac{2N}{3\pi \eta }\, \frac{H^{2}}{N^{2}}t & \text {when\, }t\ll \tau \\
 \sqrt{\frac{8}{\pi ^{3}\eta \mu _{f}\bar{\chi }^{2}}}\sqrt{\frac{H^{2}}{N^{2}}t} & \text {when\, }\tau \ll t\ll \tau N^{2}/\pi ^{2}\\
 \! \frac{2}{\pi \mu _{f}\bar{\chi }^{2}} & \text {when\, }t\gg \tau N^{2}/\pi ^{2}\end{array}
\right.\; .\]
The hydrodynamic coupling between
the particles and the confining plates was taken into account by renormalizing
the drag coefficient as  $\eta /\eta _{o}=2/[1-9a/16d]-1=1.40\; (1.7)$
for  $a=3\mu m (4 \mu m )$ \cite{FLi94}. 
 This scaling law was checked in 15 experiments where chains of $N=36$
to $59$ spheres of diameters $3$ or $4$ $\mu m$ were allowed to evolve 
from states where $\lambda$ was reduced to values between 2.7 and 267.
 In Fig.~\ref{fig3}
the average of the scaled width $WH\sqrt{a/N}$ is displayed
as a function of  the scaled time $H^{2}t/N^{2}$. 
The average  is taken both over time intervals of 0.01$t$ and  
over different experiments.

The relevance of the long-range nature of the dipolar interactions
and of the linearization of the magnetic interactions was studied via
Brownian dynamics simulations, where Eq.~(\ref{first}) without the 
$m\ddot{h}_{0}$ term was solved, this time with the full dipolar form
of the magnetic interactions computed for every particle pair.
 A repulsive  potential  $\sim exp(-100r/a)$ when $r<a$  was used to  prevent
any significant overlaps. The average width over 100
simulations was evaluated for  $N=36$, $a=3\mu m$ and  $H=4$ Oe.
 Figure \ref{fig3} demonstrates that both 
the simulations and  theoretical results agree  with 
experimental measurements in all three scaling regimes, although 
the crossover between the $t^{1/2}$ and $t^{1/4}$ regimes 
extend over a full decade. 
The   $W\sim t^{1/4}$ regime is visible over 
roughly 2.5 decades. The high variance around the average
of $W$ in the  saturated regime comes from the small number (3) of experiments
that were carried out 
 at this reduced time, along with the fact that this regime
is sensitive to the longest wavelength, for which the variance of
the amplitude is highest (it scales as $1/\omega _{n}$). %

It is also possible to obtain the dynamic scaling from 
the equilibrium behavior. For this purpose 
consider  the departure $\Delta h_i$
 from an arbitrary initial configuration, i.e. 
$\Delta h_{i}(t,t_0) = h_{i}(t+t_0 )-h_{i}(t_0)$.
Using Eq.~(\ref{eq:rel,h-n,zeta-n}) again we find that 
the averaged space Fourier transform of $\Delta h_{i}(t,t_0)$
obeys
\begin{equation}
\frac{\langle |\Delta \tilde{h}_{n}(t,t_0)|^{2}\rangle}{A t} =
\frac{1-e^{-\omega _{n}t}}{\omega
 _{n} t } \, \, ,
\label{eq:lateral,fluctu,equil}
\end{equation}
where  $A=2k_{B}T/3\pi \eta aN$.
Above we have used the saturation 
level $\langle |\tilde{h}_{n}(t_0)|^{2}\rangle =k_{B}T/\kappa N\omega _{n}$
 from Eq.~(\ref{eq:square;h-n;initial;straight}). Comparing
this expression with Eq.~(\ref{eq:square;h-n;initial;straight})
shows that the lateral fluctuations at equilibrium behave similarly
to the lateral displacements starting from a straight chain during
the non-equilibrium roughening stage, but display an amplitude $\sqrt{2}$
times larger and develop $2$ times more slowly. %
\begin{figure}
\includegraphics[  width=80in,
  height=0.80\columnwidth,
  keepaspectratio,
  angle=270,
  origin=c]{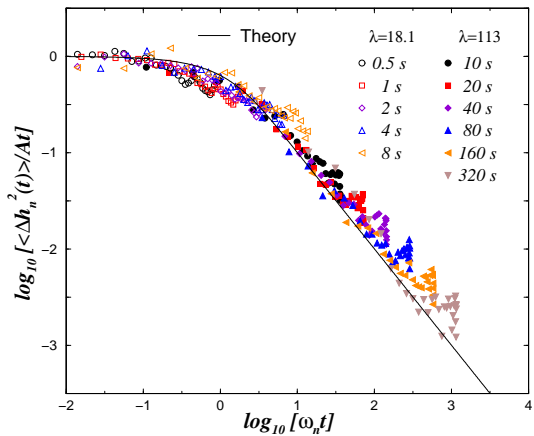}
\caption{\label{Fig:fig1}Power spectrum of the space Fourier transform of
the fluctuations normalized by time: scaling of the autocorrelation
function in time and space. The theory is given 
by Eq.~(\ref{eq:lateral,fluctu,equil}).}
\end{figure}

In  Fig.~\ref{Fig:fig1} we have compared  
the theory of Eq.~(\ref{eq:lateral,fluctu,equil}) 
to $N=57$, $a=3\mu m$ experiments with 
$\lambda =113$ and $18.1$, for which $\tau =0.022\, s$ and
$0.14\, s$. 
For each $t$ the 
power spectrum $|\Delta \tilde{h}_{n}(t,t_0)|^{2}$, calculated using a 
Hamming window, was  averaged
over all possible $t_0$'s in the 30 minutes the experiment lasted. 

The agreement between
these experiments and this theory, where there are no free parameters,
is satisfactory. Note that $\omega _{n}$ was evaluated using its
discrete form. Using the continuous asymptotic limit of the Rouse
model, $\omega _{n}=\omega _{1}n^{2}$ instead, would result in systematic
deviations from this model for large wavenumbers.

Finally, we sketch the structure of the 
connection between  the present problem and  the general
framework of anomalous diffusion of  particles in systems with
memory \cite{MOBH02}.  We consider particle $i=0$ and use Eq.~(\ref{first})
to integrate out the $i\neq 0$ variables.
By  applying time Fourier transforms and neglecting all 
but the $i=0$ mass it is possible to arrive at \cite{futurePRE}
 \begin{equation}
m\hat{\ddot{h}}_{0}(\omega )=-m\hat{\Gamma }(\omega
)\hat{\dot{h}}_{0}(\omega ) + \hat{F}(\omega )\label{eq:GLE,TFT}
\end{equation}
 where the time Fourier transform is denoted by hat symbols, 
$F(t)$ is the fluctuating part of the force and 
 $\tilde{\Gamma }(\omega )=(4/m)\sqrt{\alpha \kappa /i\omega }\, $
for $1/\tau_{1} \ll \omega \ll 1/\tau$ and $\hat{\Gamma }(\omega )\sim (4/m)\sqrt{\alpha \kappa \tau _{1}/i}$
for $\omega \ll 1/\tau _{1}$.  
The inverse Fourier transform of Eq.~(\ref{eq:GLE,TFT})
is a Generalized Langevin Equation of the Mori-Lee form \cite{MOBH02},
\begin{equation}
m\ddot{h}_{0}(t)=-m\int _{-\infty }^{t}dt_{1}\, \Gamma (t-t_{1})\dot{h}_{0}(t_{1})+F(t)\, .\label{eq:GLE}\end{equation}
From the exact form of $F(t)$ \cite{futurePRE} the fluctuation dissipation theorem for such
a generalized Langevin equation, $\left\langle F(t)F(0)\right\rangle =2mk_{B}T\Gamma (|t|)$,
can be checked directly. 
The memory effects and correlations in $F(t)$ come from the interaction
between the rest of the chain and the fluid.  According
to theory \cite{MOBH02}, the random mean square width of the observed
particle's displacement scales at long time as $W^{2}\sim t^{1-\alpha }$
if $\hat{\Gamma }(\omega )\sim \omega ^{-\alpha }$ in the limit $\omega \sim 0$.
In the present case, the asymptotic behavior of the response function
corresponds  to $\alpha =0$ for finite chains and $t > \tau _{1}$, 
and to $\alpha =1/2$ for times below $\tau _{1}$ 
($\tau_1 \rightarrow \infty$ when $N\rightarrow \infty$). 
The observed behavior
$W^{2}\sim t^{1/2}$ for $t\ll \tau _{1}$ is then in agreement with
this theory, and the present system is a simple experimental example
validating this theory.

For chains in an unbounded medium,  the Zimm model
predicts $W\simeq t^{1/3}$ \cite{Grosberg94}, and 
experiments carried on various MR fluids report
$W\simeq t^{0.35\pm 0.05}$ \cite{Furst00}. This behavior
is attributed to either  hydrodynamic-  or long rang
dipolar interactions.
In our system hydroynamic interactions are suppressed by the walls.
This may be understood by noting that the flow in our system is
described by a simple Darcy law at length scales above $d$.
This implies that a local flow perturbation decays as $1/r^2$
which is faster than the  $1/r$ decay, given by the Oseen
tensor in an unbounded medium \cite{Grosberg94}.
 Since the chain length is
typically much larger than $d$, hydrodynamic interactions are
local and do not affect the longer wavelengths of our chain
motion. On the other hand we note that if our
experimental and simulation data were 
truncated as $\log_{10} (H^{2}t/N^{2} ) \leq -1$
they could support the 
 $W\simeq t^{3/8}$ interpretation  made by Furst \& Gast \cite{Furst00}.
As these authors attribute their exponent $3/8$
to long range dipolar interactions, our results may indicate
that such interactions do play a role, but only in a crossover
regime to the $W\simeq t^{1/4}$ behavior.

In conclusion, we have established the scaling behavior of confined dipolar
chains over 5 orders  of magnitude by matching  experiments,
theory and  Brownian dynamics simulations. In passing we have made 
contact with nearby theories of both polymer models and non-Markovian 
formulations of anomalous diffusion. These results are attractive 
in particular because of the versatility and easy control of the
experiments.


\begin{thebibliography}{32}
\expandafter\ifx\csname natexlab\endcsname\relax\def\natexlab#1{#1}\fi
\expandafter\ifx\csname bibnamefont\endcsname\relax
  \def\bibnamefont#1{#1}\fi
\expandafter\ifx\csname bibfnamefont\endcsname\relax
  \def\bibfnamefont#1{#1}\fi
\expandafter\ifx\csname citenamefont\endcsname\relax
  \def\citenamefont#1{#1}\fi
\expandafter\ifx\csname url\endcsname\relax
  \def\url#1{\texttt{#1}}\fi
\expandafter\ifx\csname urlprefix\endcsname\relax\def\urlprefix{URL }\fi
\providecommand{\bibinfo}[2]{#2}
\providecommand{\eprint}[2][]{\url{#2}}

\bibitem[{\citenamefont{Skjeltorp}(1983)}]{Skj83}
\bibinfo{author}{\bibfnamefont{A.}~\bibnamefont{Skjeltorp}},
  \bibinfo{journal}{Phys. Rev. Lett.} \textbf{\bibinfo{volume}{51}},
  \bibinfo{pages}{2306} (\bibinfo{year}{1983}).

\bibitem[{\citenamefont{Toussaint et~al.}(2003)\citenamefont{Toussaint,
  Akselvoll, Flekk{\o}y, Helgesen, and Skjeltorp}}]{Toussaint03}
\bibinfo{author}{\bibfnamefont{R.}~\bibnamefont{Toussaint}} \emph{\bibinfo{author}{et al.}},
  \bibinfo{journal}{Phys. Rev. E}  (\bibinfo{year}{2003}),
  \bibinfo{note}{in press}.

\bibitem[{\citenamefont{Skjeltorp and Helgesen}(1991)}]{SH91}
\bibinfo{author}{\bibfnamefont{A.}~\bibnamefont{Skjeltorp}} \bibnamefont{and}
  \bibinfo{author}{\bibfnamefont{G.}~\bibnamefont{Helgesen}},
  \bibinfo{journal}{Physica A} \textbf{\bibinfo{volume}{176}},
  \bibinfo{pages}{37} (\bibinfo{year}{1991}).

\bibitem[{\citenamefont{Furst and Gast}(2000)}]{Furst00}
\bibinfo{author}{\bibfnamefont{E.~M.} \bibnamefont{Furst}} \bibnamefont{and}
  \bibinfo{author}{\bibfnamefont{A.~P.} \bibnamefont{Gast}},
  \bibinfo{journal}{Phys. Rev. E} \textbf{\bibinfo{volume}{62}},
  \bibinfo{pages}{6916} (\bibinfo{year}{2000}); \bibinfo{journal}{Phys. Rev. E} \textbf{\bibinfo{volume}{58}},
  \bibinfo{pages}{3372} (\bibinfo{year}{1998}); \bibinfo{author}{\bibfnamefont{S.}~\bibnamefont{Cutillas}} \bibnamefont{and}
  \bibinfo{author}{\bibfnamefont{J.}~\bibnamefont{Liu}},
  \bibinfo{journal}{Phys. Rev. E} \textbf{\bibinfo{volume}{64}},
  \bibinfo{pages}{011506} (\bibinfo{year}{2001}); \bibinfo{author}{\bibfnamefont{A.~S.} \bibnamefont{Silva}},
  \bibinfo{author}{\bibfnamefont{R.}~\bibnamefont{Bond}},
  \bibinfo{author}{\bibfnamefont{F.}~\bibnamefont{Plourbaou{\'e}}},
  \bibnamefont{and} \bibinfo{author}{\bibfnamefont{D.}~\bibnamefont{Wirtz}},
  \bibinfo{journal}{Phys. Rev. E} \textbf{\bibinfo{volume}{54}},
  \bibinfo{pages}{5502} (\bibinfo{year}{1996}).

\bibitem[{\citenamefont{Edwards and Wilkinson}(1982)}]{edwards82}
\bibinfo{author}{\bibfnamefont{S.~F.} \bibnamefont{Edwards}} \bibnamefont{and}
  \bibinfo{author}{\bibfnamefont{D.~R.} \bibnamefont{Wilkinson}},
  \bibinfo{journal}{Proc. R. Soc. London A} \textbf{\bibinfo{volume}{381}},
  \bibinfo{pages}{17} (\bibinfo{year}{1982}).


\bibitem[{\citenamefont{Grosberg and Khokhlov}(1994)}]{Grosberg94}
\bibinfo{author}{\bibfnamefont{A.~Y.} \bibnamefont{Grosberg}} \bibnamefont{and}
  \bibinfo{author}{\bibfnamefont{A.~R.} \bibnamefont{Khokhlov}},
  \emph{\bibinfo{title}{Statistical physics of macromolecules}}
  (\bibinfo{publisher}{AIP Press}, \bibinfo{address}{New York},
  \bibinfo{year}{1994}).

\bibitem[{\citenamefont{Morgado et~al.}(2002)\citenamefont{Morgado, Oliveira,
  Batrouni, and Hansen}}]{MOBH02}
\bibinfo{author}{\bibfnamefont{R.}~\bibnamefont{Morgado}},
  \bibinfo{author}{\bibfnamefont{F.~A.} \bibnamefont{Oliveira}},
  \bibinfo{author}{\bibfnamefont{G.~G.} \bibnamefont{Batrouni}},
  \bibnamefont{and} \bibinfo{author}{\bibfnamefont{A.}~\bibnamefont{Hansen}},
  \bibinfo{journal}{Phys. Rev. Lett.} \textbf{\bibinfo{volume}{89}},
  \bibinfo{pages}{100601} (\bibinfo{year}{2002}).

\bibitem[{\citenamefont{Bullough}(1996)}]{Bullough96}
\bibinfo{editor}{\bibfnamefont{W.~A.} \bibnamefont{Bullough}}, ed.,
  \emph{\bibinfo{title}{Proceedings of the 5th International Conference on
  Electro-rheological Fluids, Magneto-rheological Suspensions and Associated
  Technology}} (\bibinfo{publisher}{World Scientific},
  \bibinfo{address}{Singapore}, \bibinfo{year}{1996}).

\bibitem[{\citenamefont{Halsey and Toor}(1990{\natexlab{a}})}]{Halsey90}
\bibinfo{author}{\bibfnamefont{T.~C.} \bibnamefont{Halsey}} \bibnamefont{and}
  \bibinfo{author}{\bibfnamefont{W.}~\bibnamefont{Toor}}, \bibinfo{journal}{J.
  Stat. Phys.} \textbf{\bibinfo{volume}{61}}, \bibinfo{pages}{1257}
  (\bibinfo{year}{1990}{\natexlab{a}});  \bibinfo{journal}{Phys. Rev. Lett.} \textbf{\bibinfo{volume}{65}},
  \bibinfo{pages}{2820} (\bibinfo{year}{1990}{\natexlab{b}});
\bibinfo{author}{\bibfnamefont{T.~C.} \bibnamefont{Halsey}},
  \bibinfo{journal}{J. Colloid Interface Sci.} \textbf{\bibinfo{volume}{156}},
  \bibinfo{pages}{335} (\bibinfo{year}{1993}); \bibinfo{author}{\bibfnamefont{J.~E.} \bibnamefont{Martin}},
  \bibinfo{author}{\bibfnamefont{J.}~\bibnamefont{Odinek}}, \bibnamefont{and}
  \bibinfo{author}{\bibfnamefont{T.~C.} \bibnamefont{Halsey}},
  \bibinfo{journal}{Phys. Rev. Lett.} \textbf{\bibinfo{volume}{69}},
  \bibinfo{pages}{1524} (\bibinfo{year}{1992}).

\bibitem[{\citenamefont{Martin et~al.}(1999)\citenamefont{Martin, Hill, and
  Tigges}}]{Martin99}
\bibinfo{author}{\bibfnamefont{J.~E.} \bibnamefont{Martin}},
  \bibinfo{author}{\bibfnamefont{K.~M.} \bibnamefont{Hill}}, \bibnamefont{and}
  \bibinfo{author}{\bibfnamefont{C.~P.} \bibnamefont{Tigges}},
  \bibinfo{journal}{Phys. Rev. E} \textbf{\bibinfo{volume}{59}},
  \bibinfo{pages}{5676} (\bibinfo{year}{1999}); \bibinfo{author}{\bibfnamefont{J.}~\bibnamefont{Liu}},
  \emph{\bibinfo{author}{et al.}},
  \bibinfo{journal}{Phys. Rev. Lett.} \textbf{\bibinfo{volume}{74}},
  \bibinfo{pages}{2828} (\bibinfo{year}{1995});
\bibinfo{author}{\bibfnamefont{M.}~\bibnamefont{Fermigier}} \bibnamefont{and}
  \bibinfo{author}{\bibfnamefont{A.~P.} \bibnamefont{Gast}},
  \bibinfo{journal}{J. Colloid Interface Sci.} \textbf{\bibinfo{volume}{154}},
  \bibinfo{pages}{522} (\bibinfo{year}{1992}); 
\bibinfo{author}{\bibfnamefont{G.}~\bibnamefont{Helgesen}},
\emph{\bibinfo{author}{et al.}},
  \bibinfo{journal}{Phys. Rev. Lett.} \textbf{\bibinfo{volume}{61}},
  \bibinfo{pages}{1736} (\bibinfo{year}{1988}).

\bibitem[{\citenamefont{Fraden et~al.}(1989)\citenamefont{Fraden, Hurd, and
  Meyer}}]{Fraden89}
\bibinfo{author}{\bibfnamefont{S.}~\bibnamefont{Fraden}},
  \bibinfo{author}{\bibfnamefont{A.~J.} \bibnamefont{Hurd}}, \bibnamefont{and}
  \bibinfo{author}{\bibfnamefont{R.~B.} \bibnamefont{Meyer}},
  \bibinfo{journal}{Phys. Rev. Lett.} \textbf{\bibinfo{volume}{63}},
  \bibinfo{pages}{2373} (\bibinfo{year}{1989}); \bibinfo{author}{\bibfnamefont{J.~E.} \bibnamefont{Martin}},
  \bibinfo{author}{\bibfnamefont{J.}~\bibnamefont{Odinek}},
  \bibinfo{author}{\bibfnamefont{T.~C.} \bibnamefont{Halsey}},
  \bibnamefont{and} \bibinfo{author}{\bibfnamefont{R.}~\bibnamefont{Kamien}},
  \bibinfo{journal}{Phys. Rev. E} \textbf{\bibinfo{volume}{57}},
  \bibinfo{pages}{756} (\bibinfo{year}{1998}).


\bibitem[{\citenamefont{Cernak et~al.}(2003)\citenamefont{Cernak, Helgesen, and
  Skjeltorp}}]{Helgesen03}
\bibinfo{author}{\bibfnamefont{J.}~\bibnamefont{Cernak}},
  \bibinfo{author}{\bibfnamefont{G.}~\bibnamefont{Helgesen}}, \bibnamefont{and}
  \bibinfo{author}{\bibfnamefont{A.~T.} \bibnamefont{Skjeltorp}},
 (\bibinfo{year}{2003}),
  \bibinfo{note}{preprint}.

\bibitem[{\citenamefont{Wirtz}(1995)}]{wirtz95}
\bibinfo{author}{\bibfnamefont{D.}~\bibnamefont{Wirtz}},
  \bibinfo{journal}{Phys. Rev. Lett.} \textbf{\bibinfo{volume}{75}},
  \bibinfo{pages}{2436} (\bibinfo{year}{1995}).

\bibitem[{\citenamefont{Ugelstad and Mork}(1980)}]{Uge80}
\bibinfo{author}{\bibfnamefont{J.}~\bibnamefont{Ugelstad}} \bibnamefont{and}
  \bibinfo{author}{\bibfnamefont{P.}~\bibnamefont{Mork}},
  \bibinfo{journal}{Adv. Colloid. Int. Sci.} \textbf{\bibinfo{volume}{13}},
  \bibinfo{pages}{101} (\bibinfo{year}{1980}), \bibinfo{note}{produced under
  the trade name Dynospheres by Dyno Particles A.S., N-2001 Lillestrøm,
  Norway}.

\bibitem[{Fer()}]{Fer}
\bibinfo{note}{Type EMG 909, produced by FerroTec, 40 Simon St., Nashua, NH
  03060-3075}.

\bibitem[{\citenamefont{Bleaney and Bleaney}(1978)}]{BB78}
\bibinfo{author}{\bibfnamefont{B.}~\bibnamefont{Bleaney}} \bibnamefont{and}
  \bibinfo{author}{\bibfnamefont{B.}~\bibnamefont{Bleaney}},
  \emph{\bibinfo{title}{Electricity and Magnetism}}
  (\bibinfo{publisher}{Oxford:OUP}, \bibinfo{year}{1978}).

\bibitem[{\citenamefont{De~Gennes and Pincus}(1970)}]{DeGennes70}
\bibinfo{author}{\bibfnamefont{P.~G.} \bibnamefont{De~Gennes}}
  \bibnamefont{and} \bibinfo{author}{\bibfnamefont{P.}~\bibnamefont{Pincus}},
  \bibinfo{journal}{Phys. Kondens. Mater.} \textbf{\bibinfo{volume}{11}},
  \bibinfo{pages}{1970} (\bibinfo{year}{1970}).

\bibitem[{\citenamefont{Flekk{\o}y and Rothman}(1996)}]{flekkoy96d}
\bibinfo{author}{\bibfnamefont{E.~G.} \bibnamefont{Flekk{\o}y}}
  \bibnamefont{and} \bibinfo{author}{\bibfnamefont{D.~H.}
  \bibnamefont{Rothman}}, \bibinfo{journal}{Phys. Rev. E}
  \textbf{\bibinfo{volume}{53}}, \bibinfo{pages}{1622} (\bibinfo{year}{1996});
 \bibinfo{journal}{Phys. Rev. Lett.}
  \textbf{\bibinfo{volume}{75}}, \bibinfo{pages}{260} (\bibinfo{year}{1995}).

\bibitem[{\citenamefont{Reif}(1965)}]{reif65}
\bibinfo{author}{\bibfnamefont{F.}~\bibnamefont{Reif}},
  \emph{\bibinfo{title}{Fundamentals of statistical and thermal physics}}
  (\bibinfo{publisher}{Mc Graw-Hill}, \bibinfo{address}{Singapore},
  \bibinfo{year}{1965}).

\bibitem[{\citenamefont{Faucheux and Libchaber}(1994)}]{FLi94}
\bibinfo{author}{\bibfnamefont{L.}~\bibnamefont{Faucheux}} \bibnamefont{and}
  \bibinfo{author}{\bibfnamefont{A.}~\bibnamefont{Libchaber}},
  \bibinfo{journal}{Phys. Rev. E} \textbf{\bibinfo{volume}{49}},
  \bibinfo{pages}{5158} (\bibinfo{year}{1994}).

\bibitem[{\citenamefont{Toussaint et al.}(2003)\citenamefont{Toussaint,
Flekk{\o}y and  Helgesen}}]{futurePRE}
\bibinfo{author}{\bibfnamefont{R.}~\bibnamefont{Toussaint}},
  \bibinfo{author}{\bibfnamefont{E.~G.}~\bibnamefont{Flekk{\o}y}}, \bibnamefont{and}
  \bibinfo{author}{\bibfnamefont{G.} \bibnamefont{Helgesen}},
    (\bibinfo{year}{2003}),
  \bibinfo{note}{preprint}.

\end{thebibliography}

\end{document}